\documentclass[aip,jcp,showpacs,showkeys,reprint,eqsecnum,longbibliography]{revtex4-1}

\usepackage{amssymb}
\usepackage{amsmath}
\usepackage{amsbsy}
\usepackage{bm}
\usepackage{color}
\usepackage{graphicx}

\newcommand{\bq}{\begin{eqnarray}}
\newcommand{\eq}{\end{eqnarray}}
\newcommand{\bqn}{\begin{eqnarray*}}
\newcommand{\eqn}{\end{eqnarray*}}

\begin{document}
\title{One-dimensional fluids with positive potentials} 

\author{Riccardo Fantoni}
\email{rfantoni@ts.infn.it}
\affiliation{Universit\`a di Trieste, Dipartimento di Fisica, strada
  Costiera 11, 34151 Grignano (Trieste), Italy}

\date{\today}

\begin{abstract}
We study a class of one-dimensional classical fluids with penetrable
particles interacting through positive, purely repulsive,
pair-potentials. Starting from some lower bounds to the total
potential energy, we draw results on the thermodynamic limit of the
given model. 
\end{abstract}



\maketitle
\section{Introduction}
\label{sec:introduction}

Recently we found evidence that a non pairwise-additive interaction
fluid model for penetrable classical particles living in one-dimension
does not admit a well defined thermodynamics \cite{Fantoni2016}, but
can only exist in a zero pressure state.  

We know that physical pairwise-additive models could also have the
same thermodynamic singularity. Whereas the Ruelle stability principle
\cite{Ruelle} tells us only that a fluid of $N$ particles with a total
potential energy, $V_N$, bounded from below, $V_N>NB$ with $B$ a
constant, cannot have a divergent pressure, it does not tell us
whether it can only have a zero pressure in the thermodynamic
limit. This happens for example for models with penetrable particles
interacting with a positive, purely repulsive, long-range
pair-potential $v$.   
     
We will consider some lower bounds to the total potential energy $V_N$ 
which will allow us to prove some important results regarding the
thermodynamic limit of the underlying one-dimensional fluid model.

\section{The Problem}
\label{sec:problem}

The grand canonical partition function of a system of particles in the
segment $[0,L]$ whose positions are labeled by $x_i$ with
$i=1,2,\ldots,N$, in thermal equilibrium at a reduced temperature
$\theta$, is given by
\bq \label{omega1}
\Omega=\sum_{N=0}^\infty \frac{z^N}{N!}\int_0^Ldx_N\cdots\int_0^Ldx_1
\,e^{-\frac{V_N(x_1,\ldots,x_N)}{\theta}},
\eq
where $z>0$ is the activity. The total potential energy of the system is 
\bq \label{potential}
V_N(x_1,\ldots,x_N)&=&\sum_{i<j}v(|x_i-x_j|)\\ \nonumber
&=&\sum_{i=1}^{N-1}\sum_{j=i+1}^Nv(|x_i-x_j|),
\eq
where $v(x)$ is the pair-potential. We will assume that $v(x)\leq
v(0)=v_0<\infty$ for all $x$, {\sl i.e.} {\sl penetrable}
particles. For $v=0$ we have the ideal gas (id).   

Since $\Omega>1$ we must have for the fluid pressure $P$ 
\bq \label{P0}
&&\frac{P}{\theta}=
\lim_{L\to\infty}\frac{\ln\Omega}{L}>0,
\eq
so the pressure cannot be negative. In addition we will assume that
$v(x)$ is a {\sl positive} function, $v(x)>0$ for all $x$, then
\bq \nonumber
&&\frac{P}{\theta}=
\lim_{L\to\infty}\frac{\ln\Omega}{L}<\\ \label{P00}
&&\lim_{L\to\infty}\frac{\ln\left[\sum_{N=0}^\infty
\frac{(zL)^N}{N!}\right]}{L}=z.
\eq
So $0<P<\theta z$.

Let us furthermore assume that $v(x)$ has tails decaying to zero at
large $x$ and such that, for all $x$ in $[0,L]$, 
\bq
v(x)>v(L),
\eq
with
\bq
\lim_{L\to\infty} v(L)=0.
\eq

Then we find
\bq
\Omega<\sum_{N=0}^\infty \frac{(zL)^N}{N!}e^{-\frac{v(L)N(N-1)}{2\theta}},
\eq
and for the pressure,
\bq \nonumber
&&\frac{P}{\theta}=
\lim_{L\to\infty}\frac{\ln\Omega}{L}<\\ \label{P1}
&&\lim_{L\to\infty}\frac{\ln\left[\sum_{N=0}^\infty
    \frac{(zL)^N}{N!}e^{-\frac{v(L)N(N-1)}{2\theta}}\right]}{L}=\\ \nonumber
&&\lim_{L\to\infty}\frac{\ln\left[\int_0^\infty
dy\,\frac{(zL)^{y/\sqrt{v(L)}}}{[y/\sqrt{v(L)}]!}
e^{-\frac{y(y-\sqrt{v(L)})}{2\theta}}\right]
  -\ln[\sqrt{v(L)}]}{L},
\eq
where we introduced the new continuous variable $y=N\sqrt{v(L)}$ to
transform the series into an integral over $y$. Clearly if we had
$\lim_{L\to\infty} v(L)=v_\infty$ with $v_\infty>0$ a constant, we could
immediately conclude that the limit in Eq. (\ref{P1}) is zero (see
Eq. (\ref{L1})) and the fluid has a singular thermodynamic
limit. Since the pair-potential is defined always up to an
additive constant, in this case, in order to find a reasonable result,
one needs to properly scale the chemical potential as follows:
$\ln(z)\to\ln(z)+v_\infty (N-1)/2\theta$.  

Let us now introduce the Inverse Power Law Model (IPLM-$\alpha$),
$v(x)=v_0/[(|x|/\sigma)^\alpha+1]$, with $v_0$, $\sigma$, and $\alpha$
three positive constants, and the Generalized Exponential Model
(GEM-$\alpha$), $v(x)=v_0e^{-\gamma(|x|/\sigma)^\alpha}$, with $\gamma$
a fourth positive constant. For the IPLM-$\alpha$ with $\alpha<1$
the limit on the right hand side of Eq. (\ref{P1}) is equal to zero
(see Eq. (\ref{L4})) and the fluid can only exist in its zero pressure
state. For $1\leq\alpha<2$ it is non-zero smaller than $z$. For
$\alpha\geq 2$ it is equal to $z$ (see Eq. (\ref{L3})), {\sl i.e.} it
has the ideal gas behavior. For the GEM-$\alpha$ the limit is also
always equal to $z$. 

On the other hand we can obtain a more stringent upper bound to the
pressure observing that for models with a pair-potential with
monotonically decaying tails, {\sl i.e.} with $v^\prime(x)<0$ for all
$x$ or purely {\sl repulsive}, like the ones we just introduced, the 
configuration of minimum potential energy is approximately the one
with all particles equally spaced on the segment, so 
\bq \label{es} 
&&\min(V_N)=\\ \nonumber
&&[1+a(\alpha,N,L)]\sum_{i<j}v\left[\frac{(j-i)L}{N-1}\right]=\\ 
\nonumber 
&&[1+a(\alpha,N,L)](N-1)\sum_{k=1}^Nv\left[\frac{kL}{N-1}\right]>\\
\nonumber
&&[1+a(\alpha,N,L)](N-1)v\left(\frac{L}{N-1}\right).
\eq
For example we find, in Eq. (\ref{es}), $a(\alpha,3,L)=0$ and for
$N>3$ we generally have $a<0$. Moreover,  
\bq
\lim_{\alpha\to\infty} a=\lim_{L\to\infty} a&=&0,\\
\lim_{\alpha\to 0} a=\lim_{L\to 0} a&=&0.
\eq
Clearly $\lim_{N\to 0} a=0$ and we must also have
\bq
0<\lim_{N\to\infty}[1+a(\alpha,N,L)]\leq 1.
\eq 
So $a(\alpha,N,L)$ remains finite for all $\alpha$, $L$, and $N$ since
it must be a continuous function. 

Then we will have  
\bq \label{P3}
\frac{P}{\theta}
&<&\lim_{L\to\infty}\frac{\ln\left[\sum_{N=0}^{\infty}
\frac{(zL)^N}{N!}e^{-\frac{[1+a]\sum_{i<j}v\left[\frac{(j-i)L}{N-1}\right]}{\theta}}
\right]}{L}.
\eq
We want to study the limit on the right hand side
\bq \label{limit}
{\cal L}=\lim_{L\to\infty}\frac{\ln\left[\sum_{N=0}^{\infty}
\frac{(zL)^N}{N!}e^{-\frac{[1+a]\sum_{i<j}v\left[\frac{(j-i)L}{N-1}\right]}{\theta}}
\right]}{L}.
\eq
Now we observe that for finite $L$,
\bq \nonumber
\lim_{N\to\infty}\frac{L}{N^2}
\sum_{i<j}v\left[\frac{(j-i)L}{(N-1)}\right]&=&
\int_0^Ldx\,v(x)\\ \label{order}
&=&b(\alpha,L),
\eq
where $b(\alpha,L)$ diverges at large $L$ for the IPLM-$\alpha$ with
$\alpha\leq 1$. Then the limit of Eq. (\ref{limit}) can be easily
found for the IPLM-$\alpha$ with $\alpha\leq 1$, as   
\bq
{\cal L}=\lim_{L\to\infty}\frac{\ln\left[\sum_{N=0}^{\infty}
\frac{(zL)^N}{N!}e^{-\frac{[1+a]bL^{-1}N^2}{\theta}}\right]}{L}=0.
\eq 
So we conclude that also the IPLM-$\alpha$ with $\alpha=1$ does not
have a well defined thermodynamic limit. A pair-potential such that
$\lim_{L\to\infty}b$ is a finite constant, is said to be {\sl short
  range}.

Note that the GEM-$\alpha$ for $\alpha=1$ reduces to the Exponential
Model (EM), for $\alpha=2$ to the Gaussian Core Model (GCM), and
taking the $\alpha\to\infty$ limit of either the GEM-$\alpha$ or the 
IPLM-$\alpha$,    
\bq
\lim_{\alpha\to\infty}v(x)=\left\{\begin{array}{ll}
v_0 & |x|<\sigma\\
0   & |x|>\sigma
\end{array}\right.
\eq
we find the Penetrable Rods Model (PRM). For the PRM the
thermodynamics is well defined as follows from the analytic solution
of the Tonks gas \cite{Tonks1936} for the Hard Rods Model
(HRM). In fact we must have  
\bq
\Omega_\text{HRM}<\Omega_\text{PRM}<\Omega_\text{id}=e^{zL}.
\eq 

According to our analysis, the IPLM-$\alpha$ and the
GEM-$\alpha$ are non-singular for $\alpha\to\infty$ and the
IPLM-$\alpha$ is singular for $\alpha\leq 1$.

Moreover as already noticed in Ref. \cite{Fantoni2016} the
GEM-$\alpha$ with $\gamma \propto L^{-\alpha}$ are singular as
immediately follows from Eq. (\ref{P1}) and Eq. (\ref{L1}).

\section{External potential} 
\label{sec:ep}

In order to regularize the models introduced in the previous section,
the IPLM-$\alpha$ for $\alpha\leq 1$, which have a long-range
pair-potential, it is necessary to introduce a confining negative
external potential which will prevent the particles to ``escape'' to
infinity on the line.    

Then we will have 
\bq
V_N(x_1,\ldots,x_N)=\sum_{i<j}v(|x_i-x_j|)+N\sum_i\phi(x_i)
\eq
with $\phi$ the external potential such that $\phi(x)<-v_0/2$ for all
$x$ in $[0,L]$. So that we must now have $P/\theta>ze^{-v_0/2\theta}$.

\section{Thermodynamic regularity} 
\label{sec:tr}

In this section we want to discuss about the thermodynamic regularity
of the IPLM-$\alpha$ for $\alpha>1$, which are {\sl short-range}. We
know that $P<\theta z$. So we should look for a non-zero lower bound to
the pressure. We also know that the IPLM-$\infty$ is equivalent to the
PRM which is regular. So we can assume the IPLM-$\alpha$ to remain
regular in a neighborhood of $\alpha\to\infty$. The property that
$v(x)\leq v_0$ implies $V_N\leq N(N-1)v_0/2$ which in turn implies
$P\geq 0$ which is not enough to say that $P$ must be non-zero. 

Even if it looks plausible to assume that short-range models should
admit a regular thermodynamic limit we are unable to find a general
principle rigorously proving such an assumption.

\section{A particular non pairwise-additive model} 
\label{sec:npw}

In Ref. \cite{Fantoni2016} we studied the fluid model with
\bq
V_N&=&\sum_{i<j}w(x_i,x_j),\\
w(x_i,x_j)&=&v_0\prod_{k=i}^{j-1}A(|x_k-x_{k+1}|),\\
A(x)&=&v(x)/v_0,
\eq
where $x_1\leq x_2\leq \ldots \leq x_N$. Proceeding as in Section
\ref{sec:problem} we may assume that for equally spaced particles  
\bq \nonumber
V_N&\gtrsim& 
\text{constant}\,(N-1)\sum_{k=1}^N\left[A\left(\frac{L}{N-1}\right)\right]^k,
\eq
so that from the properties of the geometric series in the large
$N$ limit
\bq
\sum_{k=1}^N\left[A\left(\frac{L}{N-1}\right)\right]^k\sim
\frac{1-[A(L/N)]^N}{1-A(L/N)},
\eq
and choosing for $v$ the GEM-$\alpha$, this behaves as $N$ for $\alpha
> 1$ as $(1-e^{-L})N/L$ for $\alpha=1$, and as $(N/L)^\alpha$ for
$\alpha<1$. 
So from the limit in Eq. (\ref{L1}) we conclude immediately that the
model is thermodynamically singular for $\alpha>1$ with a zero
pressure, in agreement with the results of
Ref. \cite{Fantoni2016}. Nothing can be said for $\alpha\leq 1$. The
case $\alpha=1$ reduces to the physical pairwise additive model.

\section{Ensemble equivalence} 
\label{sec:ee}

In this Section we discuss the equivalence of the three thermodynamic
ensembles of statistical physics, {\sl i.e.} the grand canonical, the
canonical, and the microcanonical. The argument for the equivalence
can be found in any textbook on statistical physics, as for example
in the Course of Theoretical Physics of Landau and Lifshitz
\cite{Landau}. We briefly retrace the argument below and in the next
two subsections. 

We divide a closed system, after a period of time long enough respect
to its relaxation time, in many microscopic parts and consider one in
particular. We call $\rho(q,p)=w(E(q,p))$ the distribution function
for such part, where $q=(x_1,x_2,\ldots)$ are the particles
coordinates and $p=(p_1,p_2,\ldots)$ their momenta. In order to obtain
the probability $W(E)dE$ that the subsystem has an energy between $E$
and $E+dE$ we must multiply $w(E)$ by the number of states with
energies in this interval. We call $\Gamma(E)$ the number of states
with energies less or equal to $E$. Then the required number of states
between $E$ and $E+dE$ can be written $(d\Gamma(E)/dE)dE\propto dqdp$
and the energy probability distribution is
$W(E)=(d\Gamma(E)/dE)w(E)$. With the normalization condition $\int
W(E)dE=1$. The function $W(E)$ has a well defined maximum in
$E=\bar{E}$ and a typical width $\Delta E$ such that $W(\bar{E})\Delta
E=1$ or $w(\bar{E})\Delta\Gamma=1$, where
$\Delta\Gamma=(d\Gamma(\bar{E})/dE)\Delta E$ is the number of states 
corresponding to the energy interval $\Delta E$. This is also called
the {\sl statistical weight} of the macroscopic state of the
subsystem, and its logarithm $S=\ln\Delta\Gamma$, is called {\sl
  entropy} of the subsystem. The {\sl microcanonical} distribution
function for the closed system is
$dw\propto\delta(E-E_0)\prod_i(d\Gamma_i/dE_i)dE_i\propto
\delta(E-E_0)e^S\prod_idE_i$, where $E_0$ is 
the constant energy of the closed system and we used the fact that the
various subsystems are statistically independent so that
$\Delta\Gamma=\prod_i\Delta\Gamma_i$ and $S=\sum_iS_i$. We know that
the most probable values of the energies $E_i$ are the mean values
$\bar{E}_i$. This means that $S(E_1,E_2,\ldots)$ must have its maximum
when $E_i=\bar{E}_i$. But the $\bar{E}_i$ are the energy values of the
subsystems which corresponds to the complete statistical equilibrium
of the system. So we reach the important conclusion that the entropy
of the closed system, in a state of complete statistical equilibrium,
has its maximum value, for a given energy $E_0$ of the closed system.

\subsection{Canonical vs microcanonical} 
\label{sec:cmc}

Let us now come back to the problem of finding the distribution
function of the subsystem, {\sl i.e.} of any small macroscopic part of
the big closed system. We then apply the microcanonical distribution
to the whole system, $dw\propto\delta(E+E'-E_0)d\Gamma d\Gamma'$,
where $E,d\Gamma$ and $E',d\Gamma'$ refer to the subsystem and to the
{\sl rest} respectively, and $E_0=E+E'$. Our aim is to find the
probability $w(q,p)$ of a state of the system in such way for the
subsystem be in a well defined state (with energy $E(q,p)$), {\sl
  i.e.} in a well defined macroscopic state. We then choose
$d\Gamma=1$, pose $E=E(q,p)$ and integrate respect to $\Gamma'$,
$w(q,p)\propto\int\delta(E(q,p)+E'-E_0)d\Gamma'\propto(e^{S'})_{E'=E_0-E(q,p)}$. We
use the fact that since the subsystem is small then its energy
$E(q,p)$ will be small respect to $E_0$, $S'(E_0-E(q,p))\approx
S'(E_0)-E(q,p)dS'(E_0)/dE_0$. The derivative of the entropy respect to
the energy is just $\beta=1/\theta$ where $\theta$ is the reduced
temperature of the closed system which corresponds with that of the
subsystem in equilibrium. Then we find $w(q,p)\propto e^{-\beta
  E(q,p)}$ which is the well known {\sl canonical} distribution.     

\subsection{Grand canonical vs canonical} 
\label{sec:gcc}

We want now generalize the canonical distribution to a subsystem with
a variable number of particles. Now the distribution function will
depend both on the energy and on the number of particles $N$. The
energies $E(q,p,N)$ will be different for different values of $N$. The
probability that the subsystem contains $N$ particles and be in the
state $(q,p)$ will be $w(q,p,N)\propto
e^{S'(E_0-E(q,p,N),N_0-N)}$. Let then expand $S'$ in powers of
$E(q,p,N)$ and $N$ keeping just the linear terms, so that
$S'(E_0-E(q,p,N),N_0-N)\approx S'(E_0,N_0)-\beta E(q,p,N)+\beta\mu N$,
where the chemical potential $\mu$ and the temperature of the
subsystem and the rest are the same, since we require equilibrium. So
we obtain for the distribution function $w(q,p,N)\propto e^{\beta(\mu
  N-E(q,p,N))}$. We can define the {\sl activity} as
$z=e^{\beta\mu}$. This is the {\sl grand canonical} distribution we
chose to use throughout our discussion. 

\subsection{On the ensemble equivalence in our models} 
\label{sec:cee}

The ensemble equivalence may fail when approaching a phase transition
when the fluctuations become so large that the linear approximation
used above fails \cite{Touchette2015,Touchette2006}. This is not the
case for the models studied in the present work which do not admit a
gas-liquid phase transition since the pair-potential is lacking a
negative part (even if we cannot exclude a liquid-solid
transition). All three distribution described above, the  
microcanonical, the canonical, and the grand canonical are in
principle suitable for determining the thermodynamic properties of
our models. The only difference from this point of view lies in the
degree of mathematical convenience. In practive the microcanonical
distribution is the least convenient and is never used for this
purpose. The grand canonical distribution is usually the most
convenient. For example the Ruelle stability principle \cite{Ruelle}
holds only in this ensemble. This justifies our choice throughout the
work. 

\section{Conclusions} 
\label{sec:conclusions}

For a one-dimensional fluid model we consider some lower bounds to the
total potential energy $V_N$ which allow us to prove some results
regarding its thermodynamic limit. In particular we study fluids of
penetrable particles interacting with a positive purely repulsive
pair-potential with tails decaying to zero at infinite separations. We
study two kinds of models: The IPLM-$\alpha$ and the GEM-$\alpha$. For
the long-range models, {\sl i.e.} the IPLM-$\alpha$ for $\alpha\le 1$,
the fluid can only exist in its zero pressure state. For the
short-range models we are not able to draw any conclusion. 

We find good evidence that a particular non pairwise-additive model
already introduced in a recent previous work \cite{Fantoni2016} is
thermodynamically singular.  
   
\appendix
\section{Some limits} 
\label{sec:limit}

We have
\bq \nonumber
\lim_{L\to\infty}\frac{\ln\left[\sum_{N=0}^\infty
\frac{(zL)^N}{N!}e^{-N^{1+\epsilon}}\right]}{L}=\\ \label{L1}
\left\{\begin{array}{lc}
z   & \epsilon\leq -1\\
l   & -1<\epsilon<0\\
z/e & \epsilon=0\\
0   & \epsilon>0
\end{array}\right.
\eq
with $z/e<l<z$. For example to prove the last case $\epsilon>0$ we can
observe that 
\bq
\frac{(zL)^N}{N!}e^{-N^{1+\epsilon}}&=&\frac{(zL/e^d)^N}{N!}e^{-N(N^\epsilon-d)}\\
&<&\frac{(zL/e^d)^N}{N!},~~~\mbox{for}~~~N>d^{1/\epsilon}. 
\eq
Then for any finite $d>0$ we will find
\bq
0<\lim_{L\to\infty}\frac{\ln\left[\sum_{N=0}^\infty
\frac{(zL)^N}{N!}e^{-N^{1+\epsilon}}\right]}{L}<\frac{z}{e^d}.
\eq
Since $d$ can be chosen very large but finite, then the limit of
Eq. (\ref{L1}) must be zero. 

Also
\bq \nonumber
&&\lim_{L\to\infty}\frac{\ln\left[\sum_{N=0}^\infty
\frac{(zL)^N}{N!}e^{-N/L}\right]}{L}=\\ \label{L2}
&&\lim_{L\to\infty}ze^{-1/L}=z.
\eq

And
\bq
&&\sum_{N=0}^\infty\frac{(zL)^N}{N!}e^{-(N/L)^2}=\\
&&\sum_{N=0}^\infty\frac{(zL)^N}{N!}\sum_{k=0}^\infty (-)^k
\frac{(N/L)^{2k}}{k!}=\\
&&\sum_{k=0}^\infty(-)^k\frac{(z^2)^k}{k!}\sum_{N=0}^\infty
\frac{(zL)^{N-2k}}{N!/N^{2k}}\stackrel{L\gg\sigma}{\longrightarrow}\\
&&\sum_{k=0}^\infty(-)^k\frac{(z^2)^k}{k!}\sum_{N=2k}^\infty
\frac{(zL)^{N-2k}}{N!/N^{2k}}=e^{-z^2}e^{zL},
\eq
so
\bq \nonumber
&&\lim_{L\to\infty}\frac{\ln\left[\sum_{N=0}^\infty
\frac{(zL)^N}{N!}e^{-(N/L)^2}\right]}{L}=\\ \label{L3}
&&\lim_{L\to\infty} z-z^2/L=z.
\eq

Proceeding as above we can also prove that for the IPLM-$\alpha$ with
$\alpha>2$ and all the GEM-$\alpha$ we must have $P<\theta z$.

Moreover we have
\bq \nonumber
&&0<\frac{\ln\left[\sum_{N=0}^\infty
\frac{(zL)^N}{N!}e^{-v(L)N^2}\right]}{L}<\\ \nonumber
&&\frac{\ln\left[\sum_{N=0}^\infty
(zL)^Ne^{-v(L)N^2}\right]}{L}=\\ \nonumber
&&\frac{\ln\left[e^{[\ln(zL)]^2/4v(L)}\sum_{N=0}^\infty
e^{-v(L)[N-\ln(zL)/2v(L)]^2}\right]}{L}<\\ \nonumber
&&\frac{\ln\left[e^{[\ln(zL)]^2/4v(L)}\sum_{N=0}^\infty
e^{-v(L)N^2}\right]}{L}=\\ \label{L4}
&&\frac{[\ln(zL)]^2}{4v(L)L}+
\frac{\ln\left[\int_0^\infty dy\,e^{-y^2}\right]}{L}-
\frac{\ln[v(L)]}{2L}.
\eq
Then, since for the IPLM-$\alpha$ with $\alpha<1$ the limit of the last
expression is zero, its pressure must be zero as mentioned in the main
text. 




%

\end{document}